\documentclass[osajnl2,reprint]{revtex4-1} %% REVTeX 4.0
\usepackage{graphicx}
\usepackage{amsmath}
\usepackage{epstopdf}
\usepackage{float}
\begin{document}

\title{A versatile laser system for experiments with cold atomic gases}
\author{A. B. Deb}
\affiliation{Jack Dodd Centre for Quantum Technology,\\ Department of Physics, University of Otago, New Zealand.}
\author{A. Rakonjac}
\affiliation{Jack Dodd Centre for Quantum Technology,\\ Department of Physics, University of Otago, New Zealand.}
\author{N. Kj{\ae}rgaard}\email{nk@otago.ac.nz}
\affiliation{Jack Dodd Centre for Quantum Technology,\\ Department of Physics, University of Otago, New Zealand.}
\date{\today}

\begin{abstract}We describe a simple and compact architecture for generating all optical frequencies required for the laser cooling, state preparation and detection of atoms in an ultracold rubidium-87 experiment from a single 780~nm laser source. In particular, repump light $\sim$ 6.5 GHz away from the cooling transition is generated by using a high-bandwidth fiber-coupled electro-optic modulator (EOM) in a feedback loop configuration. The looped repump light generation scheme solves the problem of the limited power handling capabilities characterizing fiber EOMs. We demonstrate the functionality of the system by creating a high atom number magneto-optical trap (MOT).
\end{abstract}

%\ocis{000.0000, 999.9999.}% REPLACE WITH CORRECT OCIS CODES FOR YOUR ARTICLE
                          % NOTE: \ocis{} IS ALIASED TO \pacs{} BUT MUST
                          % FORMAT THE TERMS CORRECTLY FOR EACH JOURNAL

\maketitle

\section{Introduction}
Experiments with ultracold atomic species have taken a great number of directions over the last decade or so, ranging from fundamental studies of quantum physics, quantum simulation and quantum information processing \cite{mandel2003,Zoller2005} to atom optics \cite{Cronin2009}, atomic frequency standards \cite{Derevianko2011}, and ultracold chemistry \cite{Kohler2006}. As such, the experimental setups involving ultracold atoms are growing in
complexity, where more and more optical beam paths are becoming common in order to address and manipulate multiple species of atoms trapped in ultrahigh vacuum. There is thus a need for compact yet versatile laser systems for stages such as laser cooling, state preparation and state-selective detection, which are common to all these experiments. A laser system where multiple frequencies of laser light travel along the same beam path without requiring extra optical components, while offering independent and fast tunability of both the frequency and the power of each component, is highly desirable. In the present paper, we implement a laser architecture which is simple, compact, and provides all laser frequency components required for these stages with a great degree of versatility. Integrated systems such as this have attracted substantial interest recently \cite{Bongs2010, Bouyer2012, Close2012, Lienhart2007, Diao2012}.

Broadly speaking, the implementations of laser systems for cold atom experiments fall into three categories: (i) multiple lasers (some possibly boosted using optical power amplifiers), each individually frequency-locked to an atomic absorption line or a cavity resonance, (ii) multiple lasers (some possibly power boosted), one of which is frequency locked to an absorption line, while others are tied to this master laser using heterodyne beat note locks \cite{Lvovsky2012}, and (iii) a single laser frequency locked to an absorption line, the output of which is divided into parts that are manipulated with frequency shifters (acousto-optic devices) and sideband generation techniques (some of these parts possibly power boosted). The architecture presented in the current paper resides within the last category. This is based on a philosophy that it is desirable to minimize the number of optical frequency locks, which are susceptible to changes in ambient conditions. Previous implementations of this sort include direct microwave modulation of the current through a diode laser, either the master or an injection locked slave \cite{Kowalski, Myatt1993, Peng1995}. While this works well for
species with modest hyperfine splitting such as $\rm ^{85}Rb$ (3 GHz), it is not trivial to modulate a laser diode at the frequency required for $\rm ^{87}Rb$ (6.8 GHz) or $\rm ^{133}Cs$ (9.2 GHz), which are among the most popular choices for cold atom experiments. Alternatively, frequency sidebands have been imprinted externally to the laser using free space electro-optic modulators (EOMs) \cite{Streed2006, Ertmer1994}.

The recent availability of fiber optical EOMs operating at wavelengths in the 780-890 nm window, along with a desire to move in a direction towards integrated fiber optical solutions in our cold atom experiment, motivated the present work. Here we demonstrate the use of fiber optical EOMs for the generation of repump light during laser cooling despite the limited power handling capabilities of these devices. This power limitation proves to be a bottleneck when added to a linear chain of optical elements. However, by employing an optical recycling scheme, we demonstrate that sufficient repump light for operating a large MOT can be generated. Moreover, our scheme offers
the possibility of adding a repump light component to auxiliary optical probe and pumping beams at will. Our laser system capitalizes on the use of single-mode polarization-maintaining (PM) fiber optical components, such as fused fiber couplers, for splitting and combining light. This results in a compact, modular and flexible architecture.

\section{Experimental Setup}
\begin{figure}[tb!]
\centering\includegraphics[width=\columnwidth]{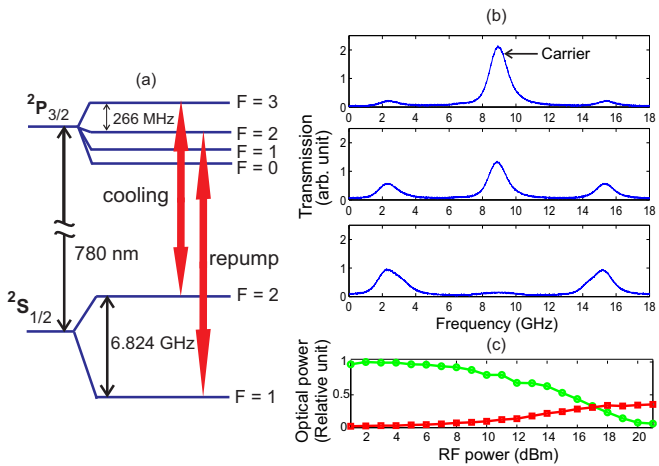}
\caption{(a) Partial atomic level structure of $^{87}$Rb with cooling and repump transitions indicated. (b) Fabry-Perot interferometer optical transmission spectra of the fiber EOM output at a driving frequency of 6.42 GHz and driving powers of 9 dBm (top), 15 dBm (center), and 21 dBm (bottom). (c) Optical power of carrier (circles) and a first-order sideband (squares) as function of rf driving power of the EOM.\label{level_sideband}}
\end{figure}
Laser cooling of $^{87}$Rb is typically performed using the cycling transition $5s ^2S_{1/2} F = 2 \leftrightarrow 5p ^2P_{3/2}  F^{'}=3 $ [see Fig.~\ref{level_sideband}(a)]. Due to off-resonant scattering from $F' = 2$, atoms can decay to the $F = 1$ state and are lost from the cooling cycle. In order to compensate for this, repumping laser light tuned to the $F = 1 \leftrightarrow F^{'} = 2$ transition is required to be present. The repump laser needs to be $\sim 6.5$ GHz higher in frequency than the cooling laser. In principle, such light can be derived from the cooling light using acousto-optic frequency shifters \cite{Buggle2000}. In practice, however, this requires a multi-pass configuration and expensive high frequency acousto-optic devices. In the present work, we have derived the repump light from the cooling light by electro-optical means, which provided both simplicity and higher efficiency. It is worth pointing out that the cost of the EOM we employ amounts to less that 15\% of the price of our commercial diode laser system which constitutes the backbone of the light sources presented in this paper. Avoiding the purchase and maintenance of multiple expensive laser systems is an important feature of the scheme presented below.

\subsection{Master laser and optical power amplifier}
The master laser for our scheme is a commercial grating-stabilized diode laser (Toptica DL-pro) with an optically isolated fiber-coupled power output of 40~mW at 780~nm. The laser is frequency-locked to the ``top-of-the-fringe'' of a $^{87}$Rb saturated absorption spectrum via feedback
control of the diode laser current and grating using a commercial FPGA-based controller (Toptica Digilock 110), making the frequency tuning highly agile and robust, enabling us to tune the laser several tens of megahertz in a millisecond.

As the optical power required for a laser cooling experiment is typically 100 mW or more in a 3D setup, we use a tapered amplifier (TA) to boost  the optical power of the master laser. Such semiconductor amplifiers offer high saturated output intensity using a structure where a single-mode ridge gain region is followed by a tapered gain region \cite{Connelly2002}. A gain of 15-20 dB is common in single-pass seeding (which we use here), whereas a gain of 37 dB has been demonstrated using a double-pass seed beam \cite{vonKlitzing2010}.

\subsection{EOM for generation of repump light}
We generate repump light using a fiber-coupled, 10~GHz bandwidth phase-modulator from Photline (NIR-MPX800). The part is a proton-exchange x-cut waveguide in a LiNbO$_{3}$ crystal \cite{Bossi2000} rated for a maximum rf power of 28 dBm , $V_{\pi} = 6.05$~V at 780~nm and a maximum optical input power of 20 mW. When applying a fixed microwave frequency drive to the EOM, a comb of red and blue sidebands are imprinted on the emitted light around the injected carrier \cite{Davis1996}. We use the first order blue sideband as a source of repump light. The calibration of relative powers for cooling (carrier) and repump (+1st sideband) using a scanning Fabry-Perot interferometer (FPI) is shown in Fig.~\ref{level_sideband}(b,c). A moderate rf power of $\sim$ 21 dBm was required to achieve an optical phase shift of $\pi$ (almost zero power at the carrier frequency in the output). The optical insertion loss (IL) limits the output to $\sim40$\% of the input power. For an input optical power of 20 mW, the maximum power in one of the maximised first-order sidebands is hence limited to about 3.8 mW, which is not sufficient to obtain large atom loading in our MOT. Moreover, the total power out of the fiber EOM (carrier and sidebands) is limited to less than 8 mW, which is not sufficient for seeding our TA.

Fiber-coupled EOMs are known to be highly sensitive to the input polarization \cite{Silander2012}. We found that a polarization extinction ratio (PER) of 20 dB or more is required for stable continuous operations at room temperature. Although the phase modulator is rated for an input optical power of 20 mW, we found that for continuous operation at this level the device is highly sensitive to the input polarization, compromising its performance. For input light with a polarization extinction ratio (PER) less than 15 dB, the output would drop to 50\% of the nominal value after only a few minutes of operation and then fluctuate over a wide range when the device is operated at room temperature. Even with light of a high PER of 26 dB, such excursions were found to occur on a characteristic timescale of an hour. We attribute this degradation in performance to a possible build-up of a photo-refractive grating inside the waveguide, where the charge carriers are moved away from the region of high light intensity, causing the output beam shape to alter and thus reducing the fiber coupling efficiency \cite{Caballero2006}.

Photorefractive damage of birefringent crystals is reversible. The usual methods involve exposing the crystal to ultraviolet light or annealing it at high enough temperatures such that the grating is erased.  We found that we can completely eliminate the EOM output fluctuations by keeping the EOM constantly operating at a temperature stabilized at 55$^{\circ}$C using a Peltier element and a temperature controller. With the device kept at this temperature, the output was stable for many hours even at relatively low PER of the input light, such as 15 dB.

We note that since the fiber EOM is rated for 780-890~nm light, we expect that the scheme described below could readily be extended to cesium MOT setups for which 852~nm light for cooling is required while  repump light is offset in frequency by 9.2~GHz (as mentioned the fiber EOM has a 10~GHz bandwidth).

\subsection{Magneto-optical trap}
We demonstrated the functionality of our scheme on our existing six-beam MOT setup \cite{Rakonjac2012} which consists of free-space optics and is fed by light from a single optical fiber carrying both cooling and repump light generated by our laser source, although ideally we would like to incorporate $\rm1\,x\,6$ fiber splitters for the six beam setup. The magnetic field gradient is generated using a pair of coils in an anti-Helmholtz configuration and is typically kept at 12 G/cm acially. We have a total of 300 mW of light available in the MOT chamber. The frequency of the cooling light is optimized to be 11 MHz red-detuned from the cycling transition. Three pairs of current-carrying coils each in a Helmholtz configuration are used to nullify stray magnetic fields at the MOT center. We determine the number of atoms in the MOT by collecting fluorescence induced by the cooling light over a solid angle and measuring the background-subtracted power on a calibrated photodiode.
\begin{figure}[tb!]
\centering\includegraphics[width=\columnwidth]{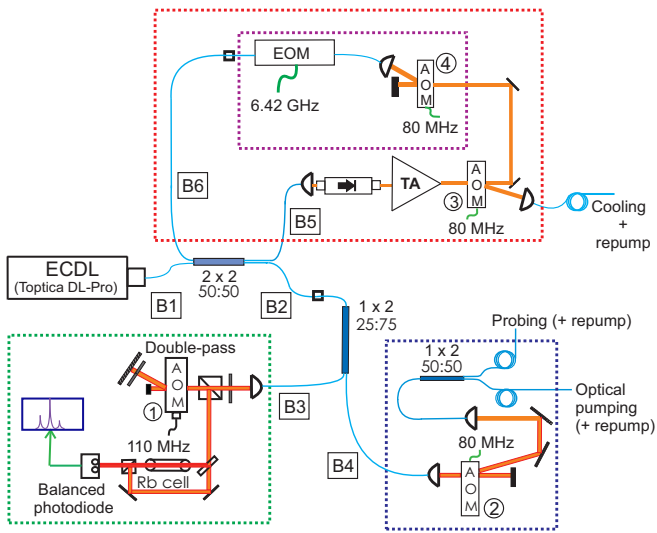}
\caption{A schematic of the laser architecture. The use of fused fiber couplers facilitate a convenient division of the system into three isolated modules, each shown in a rectangular box. See text for description.\label{scheme1}}
\end{figure}
\subsection{Feedback-loop coupling of a tapered optical amplifier}

Figure \ref{scheme1} shows a schematic of our laser source. Light from the master laser is coupled to one of the input arms (B1) of a $\rm2\,x\,2$ PM fused fiber coupler (Gooch and Housego FFP-DK5280A1P, PER 26 dB) with splitting ratio 50:50 between its outputs B2 and B5.

Light from the B2 output arm of the $\rm 2\,x\,2$ coupler is further split using a $\rm 1\,x\,2$ fused coupler with splitting ratio 25:75 (Haphit FPCL-780-751-L10-FC/APC, PER 18 dB). The light in the 25\% arm (B3) is double-passed through an acousto-optic modulator (AOM-1) with center frequency of 110 MHz and then goes through a $^{87}$Rb saturated absorption spectroscopy setup where it is locked to the $F'=2,3$ cross-over resonance. AOM-1 can be used to tune the frequency of the cooling, pumping and probing light over a range of 30 MHz. The light in the 75\% arm (B4) is brought to the vicinity of the cooling transition of $^{87}$Rb by frequency-shifting it by -80 MHz using an AOM (AOM-2). This light is then coupled to the input of $\rm 1\,x\,2$ fused coupler (Haphit FCPL-780-501-L10-FC/APC, PER 18 dB) with a 50:50 splitting ratio. The light in the output arms can be used for probing and optical pumping while AOM-2 provides fast switching of the light during such stages.

Light from the B5 output arm of the $\rm 2\,x\,2$ coupler is free-space coupled and goes through an optical isolator to a tapered amplifier chip (Eagleyard EYP-TPA-0780-01000-3006-CMT03-0000). The chip requires at least 10 mW of input light for proper seeding in a free-space configuration. The TA chip is mounted with input and output optics on a single block of metal based on a design by Nyman \textit{et al.} \cite{Nyman2006}. At 2 A of injection current and 11 mW of seed power, the TA has a gain factor of about 65 and typically puts out about 725 mW of light. The light is then frequency-shifted by -80 MHz using an acousto-optic modulator (AOM-3) to bring the frequency close to the cooling transition, which is then coupled into a fiber that carries an excess of 200 mW of light to the MOT chamber.

Due to a finite diffraction efficiency, the zeroth order of the AOM-3 has more than 100 mW of undiffracted light available, which we feed into the fiber EOM via AOM-4. This AOM controls the optical power going into the EOM and adds 80~MHz to the frequency components of the feedback light. We can put a maximum optical power of 18 mW (i.e., below the maximum rated power) into the EOM. The EOM is driven at a frequency of 6.5~GHz, which puts the blue first-order sideband of its output close in frequency to the repumping transition while the red sideband is far off resonance from any transition. Finally, the output of the EOM is fed back into the other input arm (B6) of the $\rm 2\,x\,2$ fused coupler. At 21 dBm of power driving the EOM, it can be ensured that the carrier is minimized to almost zero and the first order sidebands are maximized [see Fig.~\ref{level_sideband}(b)]. Thus, light with a repump frequency component is present on both of the output arms of the $\rm 2\,x\,2$ fused coupler. The fraction seeding the TA is amplified thus yielding sufficient repump laser power for operating the MOT. We found that a power as little as 0.65 mW at the repump frequency out of the fiber sufficed to observe a  loading of $\sim10^{10}$~atoms in the MOT. This corresponds to 9 mW of repump light in the MOT chamber which is 2.5 times higher than what we can extract from the EOM in stand-alone operation. The fraction going to the second output arm of the $\rm 2\,x\,2$ coupler can provide light at the repump transition for the optical pumping and probing stages. The fraction of light close to the repump frequency that goes into the spectroscopy module has a negligible effect when the carrier is minimized in the EOM output.
\begin{figure}[tb!]
\centering\includegraphics[width=0.75\columnwidth]{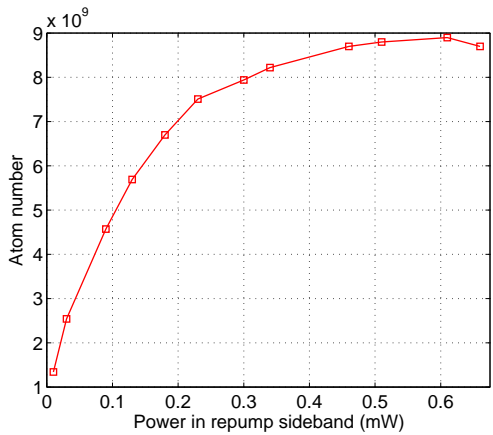}
\caption{The steady-state number of atoms in the MOT as a function of seed power into the TA at the repump sideband. The power is controlled using the acousto-optic modulator AOM-4. The total optical power used is 220 mW. The atom number could be improved further using larger diameter MOT beams as we have more power available to us.\label{scheme1MOT}}
\end{figure}

Figure~\ref{scheme1MOT} shows the steady-state number of atoms loaded in the magneto-optical trap as a function of the power at the repump frequency component in the seed beam of the TA. This power is controlled via the optical power entering the EOM by tuning the rf power driving AOM-4. The optimal level of total optical power for the MOT is found to be 220~mW, which is used for the data shown in Fig.~\ref{scheme1MOT}. A large atom number MOT, containing close to $10^{10}$ atoms, is achieved with a loading time of about 10 seconds. The MOT is highly stable requiring no adjustments of the laser source over days.

We finally point out the option of adding repump light at the outputs of the auxiliary fibers (the lower right module of Fig.~\ref{scheme1}) carrying light for optical pumping and probing. This can be done with complete control over both the power and the frequency, and opens up the possibility to prepare cold atomic samples in either of the ground hyperfine states and detect them.

\section{Conclusion and Outlook}
The atom numbers achieved in the MOTs produced by our fiber-based integrated laser architecture are comparable to typical numbers obtained in setups based on multiple diode lasers and free-space optical components that we used in the past. However, the laser system presented here is much more compact and the modular and fiber-linked approach adopted could prove very convenient for constructing portable cold atom setups \cite{Reichel2010}. We have found the systems very reliable, in particular because only one optical frequency lock is involved

In the work reported here, the repump frequency is tied to the cooling frequency such that when one shifts the latter, the absolute frequency of the former shifts accordingly. An independent control on the repump frequency is often desirable in cold atoms experiments and can be readily achieved by frequency shift keying the microwave source driving the EOM synchronously with the master laser frequency detuning. An inexpensive frequency-agile rf source suitable for this can be achieved by using a local oscillator frequency mixed with the frequency output of a direct digital synthesizer (DDS) and appropriate frequency filters \cite{Louchet-Chavet2010, Zilong2012}, where the DDS can be used to shift the frequency by up to several hundred megahertz in a fraction of a microsecond.  Along these lines, we also point out that for experiments requiring only repump light being present at times, the situtation may effectively be accomplished by detuning the (ever-present) carrier far from the resonance while keeping the repump sideband on resonance via an appropriate frequency offset.

Our complete apparatus incorporates a $^{40}$K MOT overlapped with the $^{87}$Rb MOT. We aim to combine the $^{40}$K optical pumping and probing lights with our new laser scheme for $^{87}$Rb using dual-wavelength $\rm 2\,x\,2$ fused couplers in the near future, thereby taking full advantage of the fiber-coupled apparatus. We further envisage that light consisting of highly tunable and phase coherent frequency components, as in our system, can be utilized for a STIRAP-type arrangement \cite{Vewinger2003, Oberst2007} to coherently manipulate quantum states.
\section{Acknowledgements}
{We thank Luke Taylor and Philippe LeRoux for useful discussions on fiber optical
components and fiber EOMs, respectively, and Vincent Boyer for critically reading
the manuscript. We are grateful to Sascha Hoinka for his contributions towards
constructing the MOT setup. This work was supported by Foundation for
Research, Science and Technology (FRST) Contract No. NERFUOOX0703 and Marsden Fund
Contract No. UOO1121.}

\end{document}